\documentclass[journal=mamobx, articletitle=true]{achemso}

\setkeys{acs}{articletitle = true}
\usepackage{graphicx}
\usepackage{amsmath}
\usepackage{caption}
\usepackage{subcaption}
\usepackage{amsthm,amssymb,enumerate}
\usepackage{mathtools}
\usepackage{color}
\usepackage{bm}
\usepackage{paracol}
\usepackage{changes}

\title{Influence of Topology on Rheological Properties of Polymer Ring Melts}

\author{Ranajay Datta}
\affiliation{Institute of Physics, Johannes Gutenberg University, Staudingerweg 9, 55128 Mainz, Germany}
\author{Peter Virnau}
\email{virnau@uni-mainz.de}
\affiliation{Institute of Physics, Johannes Gutenberg University, Staudingerweg 9, 55128 Mainz, Germany}

\date{\today}

\begin{document}

\begin{abstract}
We investigate with numerical simulations the influence of topology and stiffness on macroscopic rheological properties of polymer melts consisting of unknotted, knotted or concatenated rings. While melts of flexible, knotted oligomer rings tend to be significantly more viscous than their unknotted counterparts, differences vanish in a low shear rate scenario with increasing degree of polymerization. Melts of catenanes consisting of two rings on the other hand are consistently more viscous than their unconcatenated counterparts.
These topology-based differences in rheological properties can be exploited to segregate mixtures of otherwise chemically similar polymers, e.g., in microfluidic devices, which is demonstrated by exposing a blend of flexible knotted and unknotted oligomer rings to channel flow.
\end{abstract}

\section{Introduction}

\begin{figure}[ht!]
         \includegraphics[width=5.0in]{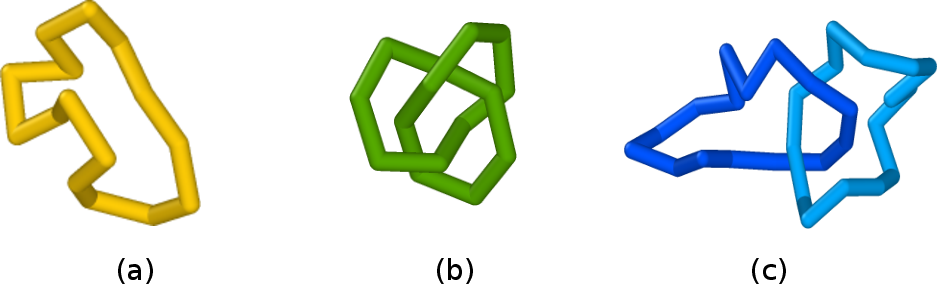}
    \caption{Ring topologies investigated in this paper: \textbf{(a)} unknotted ring polymer,   
\textbf{(b)} knotted ring polymer featuring a trefoil ($3_{1}$) knot and \textbf{(c)} pair of concatenated ring polymers. Snapshots were taken from equilibrium NVT simulations involving dense ($\rho=0.8$) melts of short ($N=15$), flexible ring polymers conducted at temperature $T=1$.}
\label{figintro}
\end{figure}

 Microscopic architecture and structural details of constituent polymers have a strong influence on dynamics and macroscopic rheological properties:
 In dilute solution, e.g., flexible star and ring polymers exhibit a weaker shear induced deformation and less pronounced shear thinning compared to linear polymers \cite{Chen_2017}.
 Suspensions of large flexible ring polymers exhibit faster diffusion, faster stress relaxation, and lower (zero shear) viscosity than systems of linear polymers of the same size \cite{Halverson_2011, Jeong_2021} for which inter-chain entanglements play a more pronounced role. Introduction of stiffness further accentuates differences in rheological properties between linear and ring polymer systems. In dense melts of semiflexible oligomer chains, an increase in chain stiffness results in emerging entanglements \cite{Faller_CPC_2001, Datta_2023, Datta_Virnau_2021, Faller1999} and a rise in viscosity \cite{Datta_Virnau_2021, Xu_2017}. At large stiffnesses, however, a nematic phase occurs which is accompanied by a drop of viscosity \cite{Datta_Virnau_2021, Xu_2017}. Dense suspensions of oligomer rings (Fig.~\ref{figintro}a) on the other hand undergo cluster formation \cite{Likos_Liebetreu_2020, Poier_2016, Bernabei_2013, Datta_2023} with increasing stiffness resulting in a steep but monotonous rise in viscosity \cite{Datta_2023}. 
 All these examples demonstrate that macroscopic rheological properties can emerge from subtle differences in the microscopic structure and interactions of polymers. In this paper we want to take this analysis one step further and investigate the influence of topology on rheology. Here, we focus on two variants, namely trefoil-knotted (Fig.~\ref{figintro}b) and concatenated ring polymers with a single link (Fig.~\ref{figintro}c), whose synthesis has received significant attention recently with the award of the Noble Prize in Chemistry in 2016 \cite{Sauvage_Nobel_2016}.

  While knots are only well-defined in closed curves \cite{Adams:1994}, physical knots in open chains can be obtained by an appropriate closure \cite{tubiana:2011:PTP}. As such
  knots can be found in various biopolymers essential for life like DNA \cite{DNA_knot_Dean_1985, Rybenkov:1993:PNAS, DNA_knot_Asuaga_2002, DNA_knot_Bao_2003, Plesa2016, Rieger:PLoS:2016, DNA_knot_Siebert_Virnau_2017, KumarSharma2019, Knots_review} and proteins \cite{Protein_knot_mansfield_1994, Taylor:Nature:2000, Protein_knot_Virnau_2006, Protein_knot_Bolinger_Virnau_2010, Protein_knot_Virnau_2011, Jarmolinska_2019, Knots_review}, and have also been subject to numerous computational studies in the context of polymer physics \cite{Vologodskii:1974, Koniaris:prl:1991, Mansfield:Macro:1994, Grosberg:PRL:2000, Virnau_JACS_2005, Marenduzzo:PNAS:2013, Trefz:PNAS:2014, Wuest:PRL:2015, Meyer_Virnau_2018, Zhao_ACSMacro_2023, Micheletti_PhRep2011, Knots_review}. 
  While the effect of inter-chain entanglements on the rheological properties of linear polymer melts has been well-documented \cite{Datta_2023, Datta_Virnau_2021, Everaers_2004, Kremer1990, KremerLee2009}, ramifications of knotting have received much less attention.
  On the one hand, the overall knotting probability in melts is actually lower than expected particularly for flexible chains \cite{Foteinopoulou_2008, Meyer_Virnau_2018, Jianrui2020, Tubiana_Virnau_2021}, and therefore it is generally assumed that knots contribute little to bulk properties of polymers \cite{Meyer_Virnau_2018}. On the other hand the transient nature of knots in linear chains and the fact there is always an equilibrium population of knots present in a melt makes it difficult to isolate their impact on bulk properties.  Here, we want to probe this important problem from a different perspective by comparing ring melts of knotted and unknotted oligomers, which circumvent these issues.

Another important variety of topological constraints are concatenations, in which two or more (typically unknotted) rings are interlocked. Concatenates have been observed in nature, e.g., in mitochondrial DNA \cite{Clayton_1967} and can occur accidentally while synthesizing ring polymers \cite{Edwards_2019}. Nowadays, they can also be produced purposefully in the lab \cite{QiongWu_2017}.
Presence of inter-ring concatethinningnations can substantially impact the dynamic properties of polymeric systems. For instance, single poly[n]catenane molecules in strong confinement exhibit slower relaxation dynamics as compared to their unconcatenated counterparts \cite{Chiarantoni_Micheletti_2}. 
 Additionally, prior simulation studies on dynamics of poly[n]catenates in dilute solution  \cite{Rauscher_2018, Rauscher_2020_2} and in the melt \cite{Rauscher_2020} have already revealed a systematic slowdown of dynamics with an increasing number of interlocked rings. An increase in stiffness of the individual rings also leads to slower dynamics\cite{Chiarantoni_Micheletti_1}. Here, we focus on the role of shear and stiffness in melts of single-concatenated oligomers.

\section{Microscopic Model and Simulation Techniques}
\label{sec_2}

 Our microscopic model depicts oligomers as bead-spring chains, as outlined in the  formulations by Kremer and Grest \cite{Kremer1990}. 
 Each bead interacts with all other beads via a repulsive Weeks-Chandler-Andersen (WCA) potential \cite{Weeks_1971}:

\begin{equation}
\begin{split}
V_{\rm WCA}(r)&=4\varepsilon\left [\left( \frac{\sigma}{r}\right) ^{12}-\left(\frac{\sigma}{r}\right)^{6} + \frac{1}{4}\right],\hspace{1cm}r\leq2^{1/6}\sigma \\
&=0,\hspace{5.3cm}r>2^{1/6}\sigma
\label{WCA}
\end{split}
    \end{equation}
where $\sigma$ is the bead diameter and $\varepsilon$ is the interaction strength, which, hereafter, will be our units of length and energy. 
Additionally, finitely extensible nonlinear elastic (FENE) bonds connect two adjacent beads:  
\begin{equation}
    V_{\rm FENE}=-\frac{1}{2}KR^{2}\ln\left[1-\left(\frac{r}{R}\right)^{2}\right]
\end{equation}
where $K$ is the spring constant and $R$ is the maximum bond extension. For the purpose of our simulations, we adopt $K=30$ and $R=1.5$ which prevent accidental chain crossing events. Stiffness is introduced in our semiflexible chains with a bending potential:

\begin{equation}
    V_\theta=\kappa(1+\cos{\theta})
    \label{bending}
\end{equation}
with $\theta$ being the angle formed by three successive beads, and $\kappa$ being the coefficient of stiffness. Such a bending potential of the cosine type owes its theoretical basis to the Kratky-Porod model. \cite{Kratky_49, DoiEdwards, RubinsteinColby} 

We apply the LAMMPS package \cite{Plimpton1995} to conduct non-equilibrium molecular dynamics simulations by imposing shear flow on dense oligomer melts at a number density $\rho=0.8$. 
LAMMPS employs a non-orthogonal simulation box featuring periodic boundary conditions, that experiences continuous deformations in compliance with external shear \cite{Evans1979, Hansen1994}, a method that has been shown \cite{Evans2008, Todd2017} to be equivalent to the Lees-Edwards boundary conditions.
The flow (f) direction aligns with the x-axis, the gradient (g) and the vorticity (v) directions are parallel to the y and z axes, respectively. 
If not stated otherwise, box sizes of $15^3$ are considered.
A constant temperature of $T=1$ is maintained with a Nos{\'e}-Hoover thermostat \cite{Evans1985, Tuckerman1997}, and the velocity Verlet algorithm is employed for integrating the equations of motion.

Shear viscosity $\eta(\dot{\gamma})$ is calculated using the relation

\begin{equation}
\eta=\frac{\sigma_{xy}}{\dot{\gamma}},
\end{equation} 
where $\dot{\gamma}$ and $\sigma_{xy}$ are applied shear rate and a non-diagonal component of the stress tensor respectively. $\sigma_{xy}$  is calculated using the Irving-Kirkwood formula \cite{Irving1950,allen-tildesley-87}:
\begin{equation}
  \sigma_{xy}=-\frac{1}{V}\left[\sum_{i}^{N}\left(m_{i}v_{i,x}v_{i,y}\right)+ \sum_{i}^{N}\sum_{j>i}^{N}\left(r_{ij,x}f_{ij,y}  \right)    \right].
\label{vis}
\end{equation}
with $\bf{v}_{i}$ and $m_{i}$ being peculiar velocity and mass of the $i^\text{th}$ particle respectively. $V$ is the volume of the system and $\bf{f}_{ij}$ and $\bf{r}_{ij}$ are the force vector and distance vector between the $i^\text{th}$ and the $j^\text{th}$ particle respectively.

 The Green-Kubo relation is utilized to estimate the zero-shear viscosity $\eta_{GK}$ to provide a reference for viscosity values at very low shear rates $\dot{\gamma}\rightarrow{0}$:

\begin{equation}
\eta_{\rm GK}=\frac{V}{k_{\rm B}T}\int_{0}^{\infty}\left\langle\sigma_{xy}(t)\sigma_{xy}(0)\right\rangle dt
\label{GK}
\end{equation}
where $k_{\rm B}$ is the Boltzmann constant.
Note that Eq.~\eqref{vis} does not consider forces originating from the thermostat and its linkage to the SLLOD conditions, which could potentially result in a slight systematic error. \cite{Jung_2016}. 

Additionally, we conduct simulations featuring a blend of ring polymers, having stiffnesses $\kappa=0$ and $10$, subject to pressure-induced channel flow by adding a constant force of $f_{x}=0.095$ in flow direction to all polymer particles. The atoms constituting the particle based channel walls are spatially arranged in an FCC formation. The particle number density of the walls is 4.0 \cite{Duan_2015}, while the particle number density of the polymer blend flowing through the channel is 0.8. Dimensions of the channel are $15\times30\times15$. Wall particles do not interact with each other and their positions are fixed throughout the simulations. Interactions between a wall particle and a polymer bead are again described by a WCA potential with diameter $\sigma_\text{wp}=0.85$  \cite{Duan_2015} and strength $\varepsilon_\text{wp}=4.0$. A DPD thermostat,\cite{Soddemann_Duenweg_2003, Pastorino_Binder_2007, Binder_DPD_2011} as implemented in LAMMPS, is used to maintain a temperature of $T=1$ across the channel. The friction coefficient pertaining to the DPD interactions, $\lambda_\text{DPD}$, is 4.5 and $r_\text{c,DPD}$. The cutoff for dissipative and random forces, is  $2\times2^{1/6}$ \cite{Pastorino_confproc_2015, Pastorino_2014} and the cutoff pertaining to WCA interactions $2^{1/6}$, as before. The structure of the channel walls, along with the interactions between the walls and polymer beads are tuned so that, upon application of a constant force \( f_{x} \) to all polymer particles in the flow direction, the resulting flow profile exhibits no-slip boundary conditions at the walls and the maximum flow velocity at the channel center.

\section{Results}

 \begin{figure}[ht!]
         \includegraphics[width=6.5in]{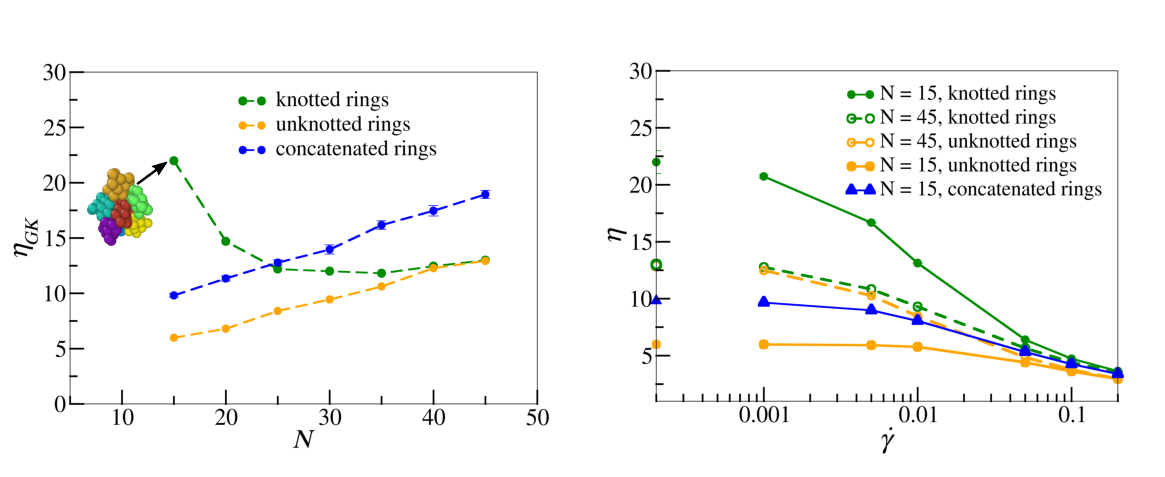}
    \caption{ \textbf{(a)} Green-Kubo viscosity $\eta_{GK}$ as function of the degree of polymerization $N$ for melts of flexible ($\kappa=0$) knotted ($3_{1}$), unknotted and concatenated ring polymers.  
\textbf{(b)}  Shear viscosity $\eta$ as a function of shear rate $\dot{\gamma}$ for melts of flexible knotted, unknotted and concatenated ring polymers with $N= 15, 45$ beads per chain. Corresponding shear viscosities according to the Green-Kubo relation $\eta_\text{GK}$ are shown on the $y$-axis. Density $\rho=0.8$ and box dimensions are $15\times15\times15$ for $N= 10, 15$ and $N\times N\times N$ for $N > 15$. Lines are guides to the eye.}
\label{fig1}
\end{figure}

Figure \ref{fig1}a displays the Green-Kubo viscosity $\eta_{GK}$ as a function of the degree of polymerization $N$ for melts of flexible knotted, unknotted and concatenated ring polymers. The unknotted ring melts exhibit a linear increase in Green-Kubo viscosity with increasing $N$\cite{Datta_2023}. On the other hand, the change in Green-Kubo viscosity with increasing $N$ for melts of knotted rings offers a rather interesting trend. At low $N$, $\eta_{GK}$ for knotted rings is significantly higher than the corresponding $\eta_{GK}$ for unknotted rings. In dense melts, as the smaller knotted rings with compact conformations and corrugated outer surfaces are closely packed against each other, the dynamics is rather slow and the viscosity is high. However, with increasing N  as the conformations become less compact, the viscosity decreases and for $N>35$ the viscosity of the knotted and the unknotted melts are almost indistinguishable. This phenomenon provides evidence that if the ring size is large enough, the presence of knots will have very little effect on the overall rheological properties in equilibrium or at low shear rates.

Figure \ref{fig1}b displays viscosity $\eta$ as a function of shear rate $\dot{\gamma}$ for melts of knotted and unknotted ring polymers with degrees of polymerization $N=15$ and $45$. While melts of  knotted and unknotted rings exhibit shear thinning for the two values of $N$, the magnitude of viscosity is polymer-size dependent only at low shear rates. At high shear rates, interestingly, while the viscosity does depend somewhat on whether the rings are knotted or unknotted, it appears to be independent of the degree of polymerization $N$. 
Zero-shear viscosity of melts of concatenated ring melts also exhibit a monotonic rise with increasing ring size, as predicted in Ref.\citenum{Rauscher_2020}. Both the Green-Kubo viscosity and the finite shear viscosity, pertaining to concatenated rings are always higher than their non-concatenated counterparts, stemming from the topological constraints imposed on the ring dynamics by the concatenations.

\begin{figure}[ht!]

 \includegraphics[width=6.5in]{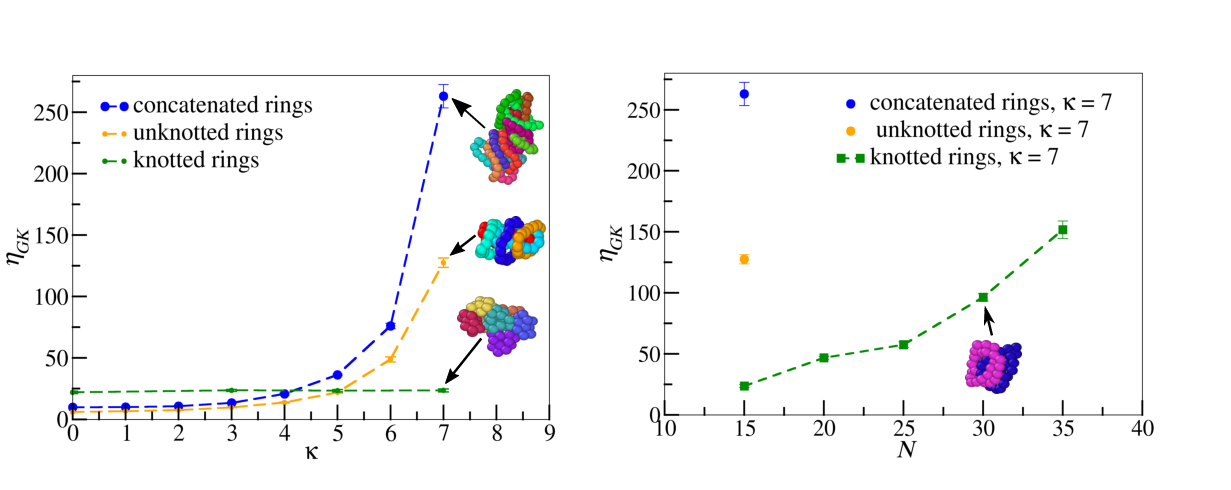}
    \caption{ \textbf{(a)} Green-Kubo viscosity $\eta_{GK}$ as a function of ring stiffness $\kappa$ for melts of concatenated, unknotted and trefoil-knotted ring polymers, corresponding to degree of polymerization, $N=15$. \textbf{(b)} Green-Kubo viscosity $\eta_{GK}$ as function of N for melts of knotted semiflexible ring polymers having  $\kappa=7$. The value of $\eta_{GK}$ corresponding to flexible ring polymers with $N=15$ and $\kappa=7$ is also shown for reference. Lines are guides to the eye. } 
\label{fig2}
\end{figure}

Melts of semiflexible knotted ring polymers also exhibit several interesting features. Earlier works investigating suspensions of semiflexible ring polymers have revealed that dense suspensions of semiflexible ring polymers exhibit the phenomenon of clustering, whereby rings interpenetrate each other to form stacks or clusters \cite{Likos_Liebetreu_2020, Datta_2023, Bernabei_2013, Slimani_2014, Poier_2015, Poier_2016} which are correlated with a slowing down of dynamics.
Interestingly, concatenated ring melts exhibit a significantly steeper increase in viscosity with increasing ring stiffness (Fig.\ref{fig2}a).
As stiffer concatenated rings also exhibit cluster formation, catenation automatically links clusters formed around the two individual rings, effectively leading to larger clusters and slower dynamics.
As expected, the rheology of small knots is almost unaffected by stiffness as the tightly wound, compact nature their conformations is unaltered (Fig.~\ref{fig2}a). For larger ring sizes (Fig.~\ref{fig2}b), however, the viscosity also increases. However, the knotted topology still prevents clustering to some degree so that viscosity is still significantly smaller than in the unknotted or concatenated case.

\begin{figure}[ht!]
     
      \includegraphics[width=6.2in]{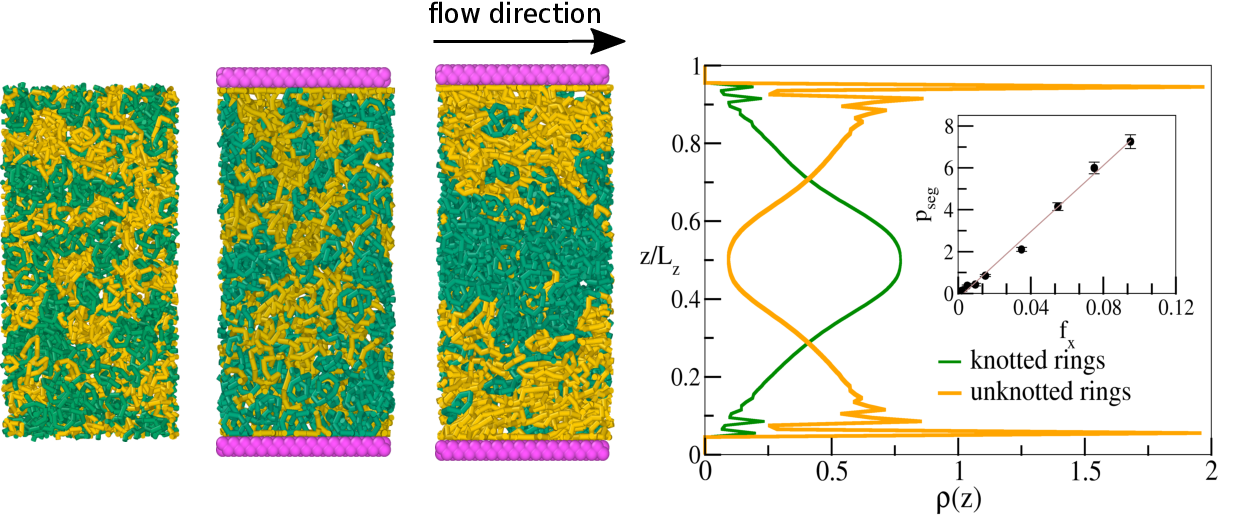} 
        \caption{\textbf{(a)} Snapshot of a homogeneous binary mixture of unknotted flexible ($\kappa=0$) ring polymers (yellow) and flexible trefoil-knotted (green) rings. The blend is equimolar, i.e., the fraction of knotted rings is $\chi_k=0.5$. \textbf{(b)} The same mixture confined between two layers of particle-based walls (shown in pink). In equilibrium, the two components of the mixture do not segregate, even in the presence of confining walls. \textbf{(c)} The confined mixture is subjected to channel flow (by the application of a constant force of $f_{x}=0.095$ along the $x$-axis to all the polymer particles) and exhibits segregation. A box size of $15\times15\times30$ was considered for part (a) and the channel size was $15\times15\times30$ for parts (b) and (c). \textbf{(d)} Density profiles of the respective components along the channel cross-section. The inset shows the plot of the segregation parameter, \( p_{seg} = \left( \frac{\rho_{knot}}{\rho_{unknot}} \right) - 1 \) as a function of the applied force $f_{x}$. Here, \( \rho_{knot} \) and \( \rho_{unknot} \) refer to the monomer number densities of knotted and unknotted rings, respectively, at the channel center. A linear fit to the data points is shown in brown. }
        \label{snapshots}
    \end{figure}

In the final section of our work, we study flow-induced segregation in an equimolar binary blend of flexible knotted and unknotted oligomer ($N=15$) rings at an overall density of $\rho=0.8$ and temperature $T=1$. In bulk and in confinement between two walls, knotted and unknotted rings form a homogeneous mixture as indicated in Figs.~\ref{snapshots}a,b. 

 When the confined mixture is subjected to channel flow by applying a constant force $f_x=0.095$ in flow direction to all beads, knotted and unknotted rings segregate (Fig.~\ref{snapshots}c). While the former accumulate in the central region of the channel where the shear rate is lowest, unknotted rings migrate towards the channel walls where the shear rate is highest.
Fig.~\ref{snapshots}d displays the density profiles of the two components under flow and clearly establishes flow-induced segregation.

To quantify the extent or degree of phase segregation as a function of the applied force \( f_x \), we introduce the parameter \( p_{seg} = \left( \frac{\rho_{knot}}{\rho_{unknot}} \right) - 1 \), where \( \rho_{knot} \) and \( \rho_{unknot} \) represent the monomer number densities of knotted and unknotted rings, respectively, at the channel center. By this definition, larger values of \( p_{seg} \) indicate greater degrees phase segregation. As shown in the inset of Fig.~\ref{snapshots}d, \( p_{seg} \) increases continuously, linearly, and monotonically with increasing \( f_x \), indicating a progressive increase in 
segregation with $f_{x}$. 
 
\begin{figure}[ht!]

      \includegraphics[width=6.6in]{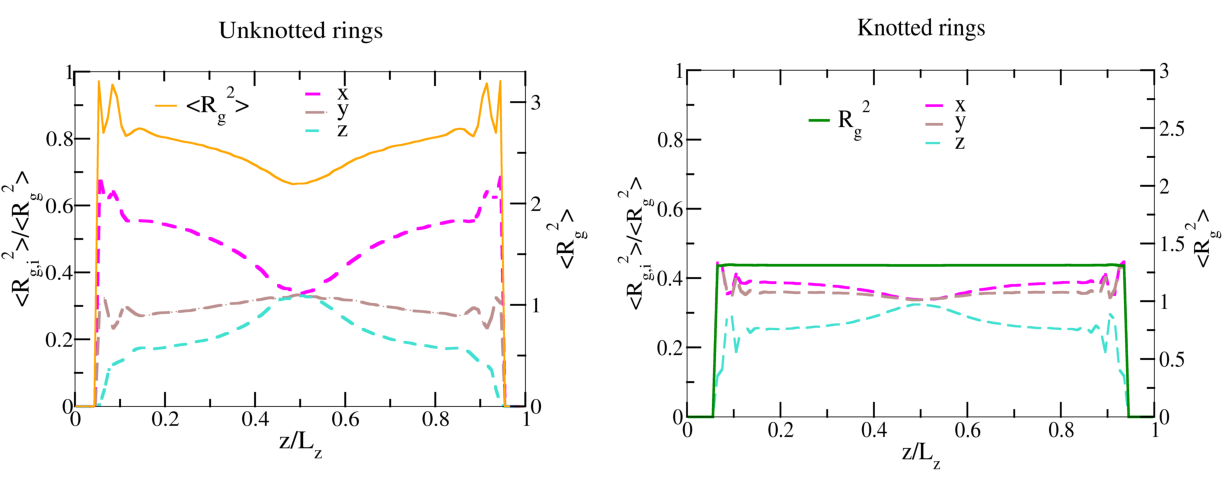}

\caption{ $\langle R_{\text{g},i}^2 \rangle / \langle R_\text{g}^2\rangle$ profiles across the channel cross section, where $i$ is a placeholder for x,y  and z directions for the unknotted (\textbf{a}) and knotted (\textbf{b}) rings. $\langle R_\text{g}^2\rangle$ profiles are also shown.
Note that in our setup, $f_{x}$ is applied along the $x$ direction. Hence $x$ is the flow (f) direction and $z$ and $y$ are the gradient (g) and vorticity (v) directions, respectively. }
\label{orientations_and)density_profiles}
\end{figure}

Figure \ref{orientations_and)density_profiles}a demonstrates that the unknotted rings undergo appreciable stretching and orientation in the direction of shear in the high shear rate regions near the walls while the knotted rings, with their tightly wound compact conformations, hardly adapt (Fig.~\ref{orientations_and)density_profiles}b).
Just like in previous work on mixtures of star and chain polymers \cite{Srivastava_Nikoubashman_2018}, ring and chain polymers  \cite{Weiss_Nikoubashman_2019} and recently on a blend of flexible and semi-flexible oligomer rings \cite{Datta_2023} we observe a tendency of the less deformable component to migrate towards regions of low shear (i.e., the center) where deformation forces are smallest, while the more deformable (and thus adjustable) component is driven into regions of high shear.

\section{Conclusions and Outlook}
In this study we have investigated the impact of topology of individual constituent ring polymers on rheological properties of melts. 
By choosing two simple model systems, melts comprising entirely of trefoil knots and concatenated rings, we provide an initial assessment of how self-entanglements influence rheological properties of polymer melts.

Our numerical results show that melts of small knotted ring polymers have significantly elevated viscosities compared to melts of unknotted rings. However, this effect vanishes as the degree of polymerization increases. 
Based on these results and given that knotting probabilities are typically low in melts of long linear polymers \cite{Meyer_Virnau_2018}, we conject that unlike entanglements, knots do not contribute significantly to the rheology of such melts in line with previous observations \cite{Kapnistos2008}.
 At higher shear rates, however, viscosity differences between knotted and unknotted ring melts persist to some degree, indicating a complex interplay between topology and shear-induced deformation. The importance of topology is further illustrated by our investigation of melts of single concatenated rings. The latter always exhibit a larger viscosity than their non-concatenated counterparts, particularly as chain stiffness increases.

Finally, we observe that while an equimolar blend of small, flexible knotted and unknotted rings are homogeneously mixed in equilibrium, it segregates when exposed to channel flow. This finding offers potential applications in microfluidic devices for the separation of polymers that are similar in mass and degree of polymerization, but differ in viscoelastic (and rheological) properties emerging from topology.





\newpage
\bibliography{cit1}

\end{document}